\documentclass[prb,twocolumn,showpacs,superscriptaddress,amsmath,floatfix]{revtex4-2}
\usepackage{xr-hyper} 
\usepackage[colorlinks=true, linkcolor=blue, citecolor=blue, urlcolor=blue]{hyperref}

\makeatletter
\newcommand*{\addFileLabels}[2][]{%
  \let\org@theH@internal\theH@internal%
  \define@key{Gm}{#1}{}%
  \externaldocument[#1]{#2}%
}
\makeatother

\addFileLabels[supp-]{supplement}

\usepackage{amssymb}
\usepackage{color}
\usepackage{dcolumn}
\usepackage{graphics,psfrag}
\usepackage{ulem}
\usepackage{float}
\usepackage{gensymb}
\usepackage{graphicx}
\usepackage{makecell}
\usepackage{mathtools}
\usepackage{xcolor}
\setcitestyle{square,numbers}
\usepackage{appendix}

\definecolor{deep-blue}{rgb}{0.17, 0.17, 0.89} 
\def \tc {$T_c$}
\def \chiper {$\chi_{\perp}^{-1}$}
\def \chipar {$\chi_{\parallel}^{-1}$}

\begin{document}

\title{Thermally melted quadrupolar order and intrinsically quantum phases in 5$d^1$ double perovskites}
\author{Rong Cong}
\thanks{Current affiliation: NHMFL, FL}
\thanks{Equal contribution}
\affiliation{Department of Physics, Brown University, Providence, Rhode Island 02912, USA}
\author{Ginevra Corsale}
\thanks{Equal contribution}
\affiliation{Department of Physics, Brown University, Providence, Rhode Island 02912, USA}
\author{Ilija K. Nikolov}
\affiliation{Department of Physics, Brown University, Providence, Rhode Island 02912, USA}
\author{Wenjuan Zhang}
\affiliation{Department of Physics, Ohio State University, 191 West Woodruff Ave
Columbus, Ohio 43210, USA}
\author{Nandini Trivedi}
\affiliation{Department of Physics, Ohio State University, 191 West Woodruff Ave
Columbus, Ohio 43210, USA}
\author{Vesna F.  Mitrovi\'c}
\email[Corresponding author: ]{vemi@brown.edu}
\affiliation{Department of Physics, Brown University, Providence, Rhode Island 02912, USA}
\affiliation{Brown Center for Theoretical Physics and Innovation, BCTPI, Brown University, Providence, Rhode Island 02912-1843, USA}

\date{\today}
\begin{abstract}
\textbf{Abstract.} 
Spin-orbit-coupled $d^1$ double perovskites exhibit a rich interplay of spin, orbital, and quadrupolar degrees of freedom (DOF), giving rise to competing magnetic and multipolar phases. Although quantum mean-field theories predict many exotic phases, their stability against thermal fluctuations and reproducibility within a classical framework remain an open question. Here, we surpass the mean-field limitations by deploying large-scale classical Monte Carlo simulations on the projected $j=3/2$ manifold of the FCC lattice, allowing complex ordering patterns to emerge spontaneously without preassigned magnetic symmetries. Our thermodynamic mapping reveals that thermal fluctuations melt the
intermediate-temperature quadrupolar phase over part of the phase diagram, while it survives intact elsewhere. Crucially, by systematically isolating the boundary between classical and quantum stability, we demonstrate that while the four-sublattice antiferromagnetic and ferromagnetic (FM) phases are robustly classical, the coplanar canted $\text{FM}[110]$ state completely destabilizes. This identifies the $\text{FM}[110]$ phase as an intrinsically quantum state born out of quantum fluctuations. Our results demonstrate that the dominant magnetic phases are robust within a classical description, where the essential physics of the system is captured by weakly entangled, short-range correlated DOF and highlight the role of thermal fluctuations in determining the stability of different types of magnetic and quadrupolar order.
\end{abstract}
\maketitle

\section*{Introduction}
Transition metal oxides (TMOs) exhibit a wide variety of electronic and magnetic phenomena, including colossal magnetoresistance, multiferroicity, and high-temperature superconductivity~\cite{pesin2010mott, khomskii_orbital_2021, Khomskii2014, Chen:2024aa, Khomskii2024}.
In $5d$ TMOs, this complexity is heightened by the strong spin-orbit coupling~(SOC) intertwined with other interactions, such as Hund's coupling and crystal field splitting~\cite{chen2009spin,chen2010exotic,chen2011spin_d2,jackeli2009mott,radic2012exotic,cole2012bose,reuther2011finite,nussinov2015compass, rau2016spin, Svoboda2017, Svoboda2021,Iwahara_2024}. Together, they produce rare quantum phases, such as multipolar magnetic order~\cite{witczak2014correlated, MonteCarlo_2014magnetism}, axion insulators~\cite{wan2012computational}, quantum spin liquids~\cite{okamoto2007spin,lee2008end,balents2010spin, Carvalho2023}, which are often undetectable in many standard probes and thus known as hidden order~\cite{Carr2022, Voleti2023, NikolovSS23, pourovskii2025hidden}. 

The $5d^1$ A$_2$BB'O$_6$ double perovskites have emerged as an important platform for studying multiflavor Mott physics~\cite{Chen:2024aa, Frontini2024}. The inherent geometric frustration of the face-centered cubic (FCC) lattice cooperates with strong SOC to stabilize a rich landscape of exotic magnetic and multipolar phases~\cite{Marjerrison2016, Romhanyi2017, hirai2019successive, Hirai2020, Streltsov2020,ishikawa2021phase,CruzPinhaBarbosa2022, Streltsov2022, DaCruzPinhaBarbosa2024, Zivkovic2024, Soh2024, FioreMosca2024, FioreMosca2024a, Sutcliffe2026, Nikolov2026emergence}. In particular, the double-perovskite osmates $\text{Ba}_2M\text{OsO}_6$ ($M = \text{Li, Na}$) offer a compelling demonstration of the delicate balance among competing ground states~\cite{stitzer_crystal_2002, Erickson2007, Steele2011}. The sodium variant, $\text{Ba}_2\text{NaOsO}_6$, hosts a glassy phase~\cite{Nikolov2026} that precedes exotic canted antiferromagnetism (cAFM)~\cite{Lu2017, Liu2018, Liu2018a, Willa2019, Cong2019, Cong2020, Cong2023, Agrestini2024}, further tuned via calcium doping ($\text{Ba}_2\text{Na}_{1-x}\text{Ca}_x\text{OsO}_6$) into a collinear AFM phase affected by bipolarons~\cite{Kesavan2020, Cong2023, Celiberti2024}. Replacing sodium with lithium ($\text{Ba}_2\text{LiOsO}_6$) seems to drastically change the structural flexibility, yielding simpler AFM order~\cite{stitzer_crystal_2002, Erickson2007}.

The mounting evidence for exotic multipolar ground states has driven the development of microscopic models capturing the competition between dipolar and quadrupolar interactions under SOC. Early theoretical work, based on effective Hamiltonians in the strong SOC limit by Chen \textit{et al.}~\cite{chen2010exotic} and subsequently extended by Svoboda \textit{et al.}~\cite{Svoboda2021} to explicitly account for SOC and $j=3/2$, $j=1/2$ mixing, identified a broad collection of candidate ground states. However, as these studies rely on mean-field approximations, it remains unclear how inter-site correlations affect the stability of the predicted phases.

Here, we extend the projected Hamiltonian of Chen \textit{et al.} beyond mean-field theory using large-scale classical Monte Carlo simulations. Simulating extended lattices without imposing predetermined magnetic structures allows energetically preferred ordering patterns to emerge spontaneously, capturing both long-range spatial correlations and thermal fluctuations. Our calculations reproduce the AFM 4-sub and FM 4-sub phases as the dominant low-temperature ordered states while revealing significant changes in the finite-temperature phase diagram invisible to mean-field approximations. We find that the AFM 4-sub phase emerges through a single transition from the paramagnetic (PM) state in the regime dominated by antiferromagnetic exchange. As the ferromagnetic exchange and quadrupolar interactions increase, the FM 4-sub phase becomes the stable low-temperature state, reached either directly from the PM phase or through intermediate ordered phases depending on the relative strength of the competing interactions. In particular, we identify both intermediate quadrupolar and AFM 4-sub regimes preceding the FM 4-sub state, as well as a region where the quadrupolar phase is suppressed and the system undergoes a direct PM-FM 4-sub transition. Finally, a direct comparison with prior quantum mean-field predictions~\cite{chen2010exotic, Svoboda2021} maps the boundary between effects originating from classical and quantum dynamics.

\section*{Results}

\subsection*{Model and simulation}
\begin{figure}[t]
    \centering
    \includegraphics[width=1\linewidth]{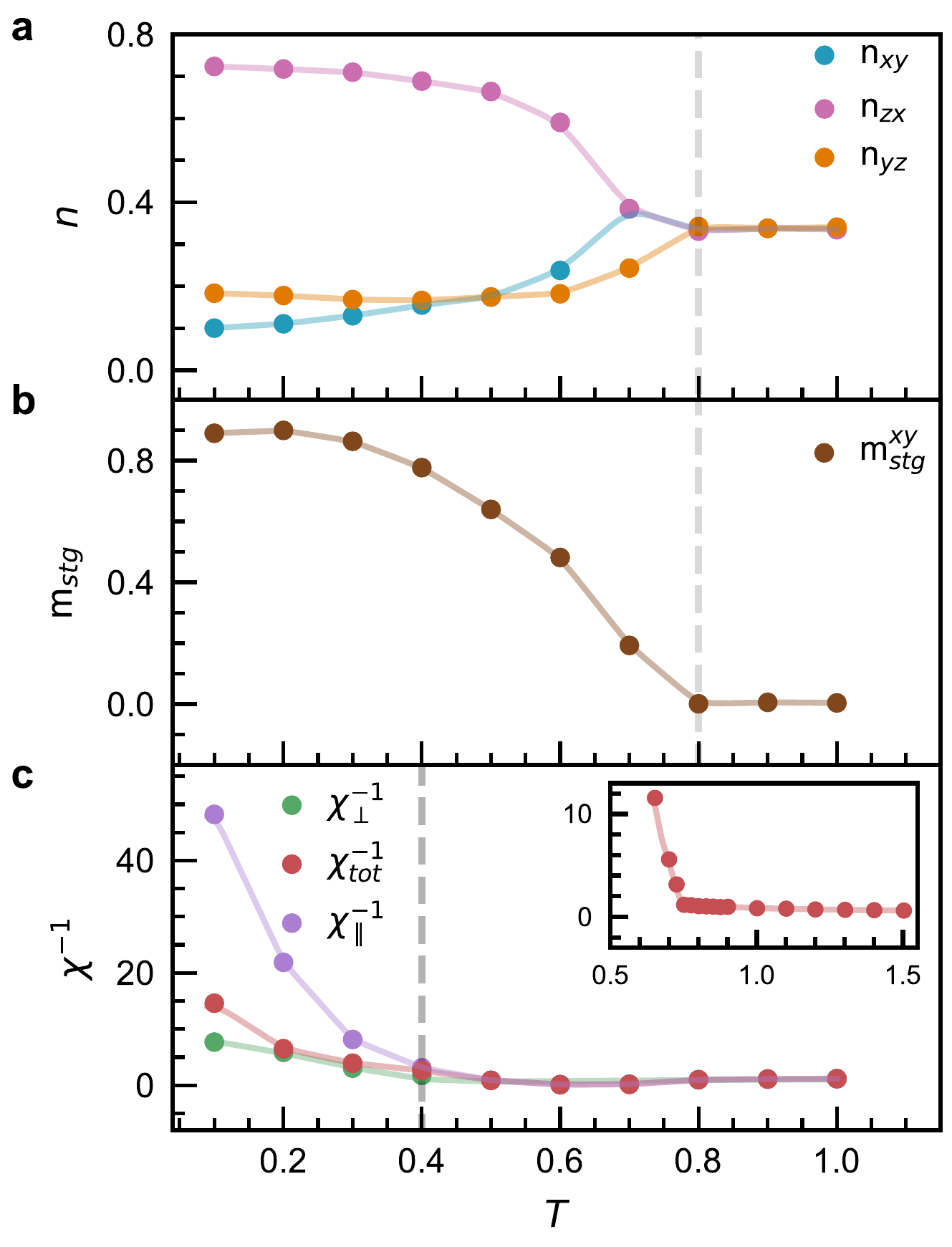}
    \caption{\textbf{Pure FM coupling.}
    \textbf{a} Orbital occupations $\langle\tilde{n}_{\alpha\beta}\rangle$ vs. temperature, show cubic symmetry breaking and orbital polarization below $T_c$. 
    \textbf{b} Staggered magnetization $m_{\mathrm{stg}}^{xy}$ developing continuously with $T_c \sim 0.8$ of the PM to FM 4-sub transition. 
    \textbf{c} Total ($\chi^{-1}$), longitudinal ($\chi_{\parallel}^{-1}$), and transverse ($\chi_{\perp}^{-1}$) inverse susceptibilities relative to the ordered moment, displaying strong anisotropic splitting below $T\sim0.4$; $\chi_{\parallel}^{-1}$ reflects suppressed longitudinal fluctuations, whereas small $\chi_{\perp}^{-1}$ indicates soft transverse modes. Inset: Deviation from CW behavior yields a negative $\Theta_{\mathrm{CW}}$. 
    Gray dashed lines mark the transition temperature.
    Results shown for $12^3\times4$ lattices. 
    }
    \label{fig:pure_fm}
\end{figure}

We adopt the microscopic model introduced by Chen \textit{et al.}~\cite{chen2010exotic} for a $d^1$ double-perovskite system. The Hamiltonian incorporates nearest-neighbor antiferromagnetic~(AFM) exchange~$(J_1)$,  nearest-neighbor ferromagnetic~(FM) exchange~$(J_2)$, electric quadrupole-quadrupole interactions~$(V)$, and SOC~$(\lambda)$. The full mathematical description is provided in the Methods. In the strong SOC limit~$(\lambda~\rightarrow~\infty)$, the spin and orbital degrees of freedom (DOF) become strongly entangled, and the Hamiltonian is projected onto the $j=3/2$ basis. Since the system contains a single $d$ electron per magnetic site, the three $t_{2g}$ orbitals ($xy$, $yz$, and $zx$) satisfy the single-occupancy constraint of Eq.~\eqref{occupation_number1}. When projected onto the $j=3/2$ basis, the $t_{2g}$ constraint relates the orbital occupations to quadratic combinations of the angular momentum components, as detailed in the Methods. 

To characterize the magnetic and quadrupolar phases together with their thermal response, we perform classical Monte Carlo (MC) simulations in which the effective moment $\mathbf{j}_i$ is represented by a classical vector of fixed length $j=\sqrt{15}/2$, rather than by a quantum operator. The single-occupancy constraint preserves the local structure of the quantum $j=3/2$ manifold within the classical representation. Specifically, it restricts the allowed orientations of  $\mathbf{j}_i$ to configurations that correspond to a valid occupation of the $t_{2g}$ orbitals. As a result, the classical moments do not explore the full range of configurations available to an unconstrained classical spin, but remain restricted to the subset consistent with the projected $j=3/2$ quantum description.

The simulation utilizes a standard Metropolis algorithm on the B’ magnetic sites forming an FCC lattice with four sites per unit cell, corresponding to the four sublattices (4-sub) used to describe the ordered phases. We simulate system sizes ranging from $6^3\times4$ to $12^3\times4$ sites, with the system slowly annealed from high temperature to low temperature to obtain the equilibrium behavior across the phase diagram. A finite-size analysis shows that the qualitative thermal behavior is unchanged with increasing system size, while the main effect of larger lattices is a progressive sharpening of the transition features due to reduced finite-size effects (see Supplementary Information). Detailed definitions of the computed order parameters and observables are provided in the Methods.

Table~\ref{tab:1} summarizes the principal methodological differences between the present work and the mean-field studies of Chen \textit{et al.}~\cite{chen2010exotic} and Svoboda \textit{et al.}~\cite{Svoboda2021}. Although all three approaches investigate the same microscopic model, they differ in the treatment of SOC, the representation of the local DOF, and the computational framework.

To identify the phases favored by the different competing interactions, we first 
examine the limiting cases in which only one term of the Hamiltonian is present. The simplified limits serve as a guide for interpreting the behavior of the full model, where all interactions compete simultaneously.

\subsection*{Pure antiferromagnetic phase $(J_1=1)$}\label{4afm}
\begin{figure}[t]
    \centering
    \includegraphics[width=1\linewidth]{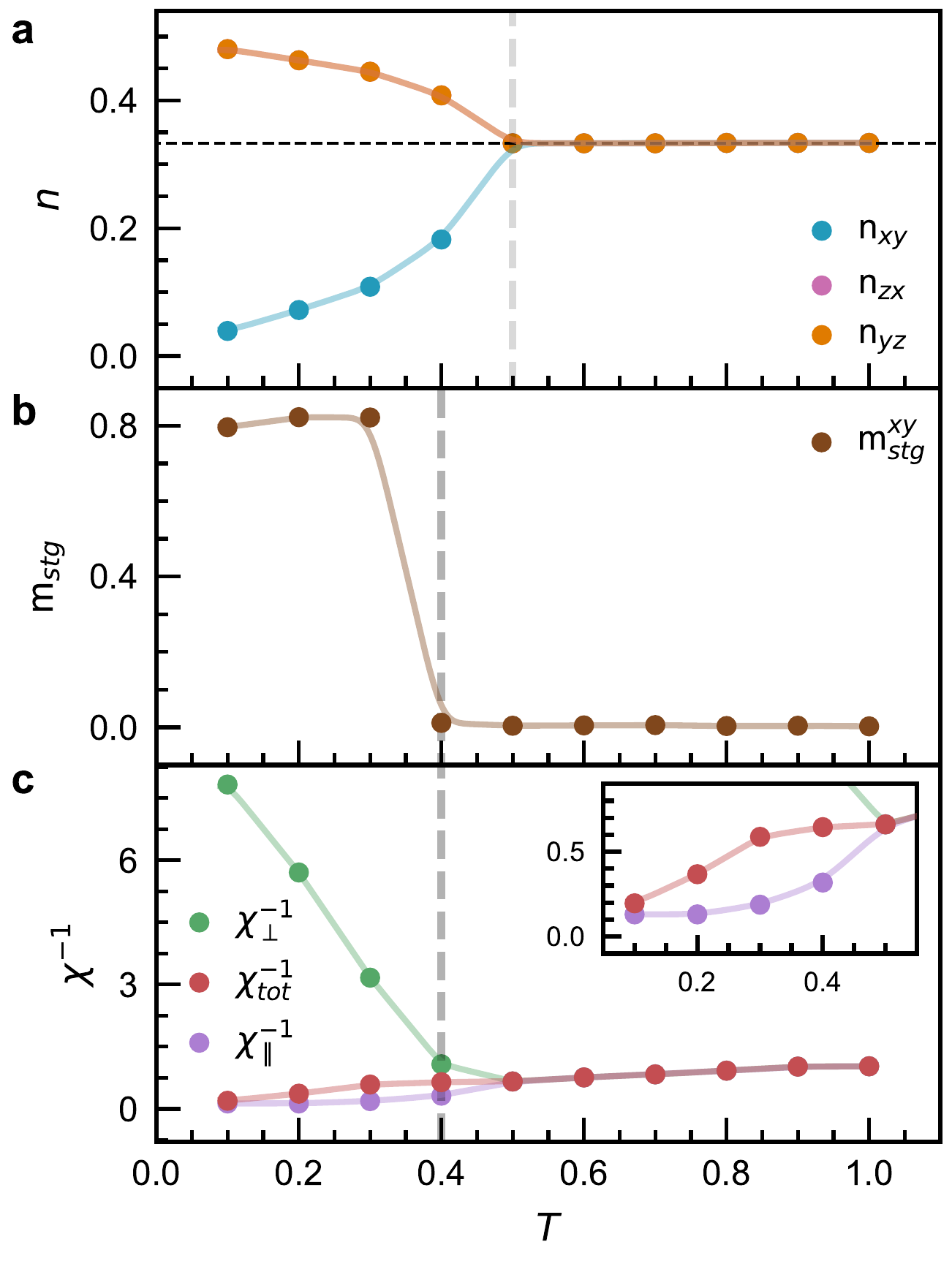}
    \caption{\textbf{Pure AFM coupling.}
    \textbf{a} Orbital occupations $\langle\tilde{n}_{\alpha}\rangle$ vs. temperature show spontaneous cubic symmetry breaking below $T \sim 0.5$ ($\langle\tilde{n}_{yz}\rangle=\langle\tilde{n}_{zx}\rangle>\langle\tilde{n}_{xy}\rangle$).
    \textbf{b} Staggered magnetization $m_{\mathrm{stg}}^{xy}$ marking the transition to the AFM 4-sub phase at $T_c \sim 0.4$.
    \textbf{c} Total ($\chi_{\mathrm{tot}}^{-1}$), parallel ($\chi_{\parallel}^{-1}$), and perpendicular ($\chi_{\perp}^{-1}$) inverse susceptibilities relative to the $z$ axis, displaying strong anisotropy below $T_c$. Below $T_c$, $\chi_{\perp}^{-1}$ increases as in-plane magnetic fluctuations freeze, whereas small $\chi_{\parallel}^{-1}$ indicates strong $z$-axis susceptibility. Concurrent magnetic and orbital ordering highlights spin-orbital coupling.
    Inset: Non-linear deviations from PM behavior. 
    Gray dashed lines mark the transition temperature.    
    Results shown for $12^3\times4$ lattices.
    }
    \label{fig:pure afm}
\end{figure}
When only the AFM interaction is present, the system stabilizes into the AFM 4-sub phase upon cooling. This state is characterized by four magnetic sublattices, with antiparallel in-plane moments within each layer and a 90$\degree$ rotation between neighboring layers, illustrated in Fig.~S1b of the Suppl. The resulting structure has a vanishing uniform magnetization, while the staggered in-plane magnetization captures the underlying AFM order.
The temperature evolution of the orbital occupations is shown in Fig.~\ref{fig:pure afm}a. Below the transition temperature, the orbital occupations satisfy  $\langle\tilde{n}_{yz}\rangle = \langle\tilde{n}_{zx}\rangle>\langle\tilde{n}_{xy}\rangle$ consistent with previous mean-field results~\cite{chen2010exotic}.
The orbital imbalance reflects the spontaneous breaking of cubic symmetry associated with the AFM 4-sub order. Through the projected orbital operators in Eq.~\eqref{new operators}, the orbital occupations are directly related to the orientation of the effective moment $\mathbf{j}$. The predominantly in-plane ordered moments therefore lead to the observed anisotropic orbital distribution.

The magnetic transition is characterized by the development of the staggered magnetization $m_{stg}^{xy}$, shown in Fig.~\ref{fig:pure afm}b, becoming finite below $T_c\sim0.4$. While the uniform magnetization remains zero, $m_{stg}^{xy}$ becomes finite below $T_c\sim0.4$, signaling the onset of AFM order. The transition is accompanied by the splitting of the orbital occupations, indicating that magnetic and orbital ordering emerge simultaneously. The inverse susceptibility (Fig.~\ref{fig:pure afm}c) reveals a strong anisotropy between the $z$ direction and the $xy$ plane (Fig.~\ref{fig:pure afm}c), reversing the trend observed in the FM-only case. Below \tc, \chiper rises sharply, indicating that fluctuations within the $xy$ plane are suppressed as the staggered moments develop, while \chipar remains comparatively small, corresponding to a larger susceptibility along the $z$ direction. Such anisotropy is consistent with an AFM state whose ordered moments lie within the $xy$ plane. The above scenario is further confirmed by the real-space correlation functions as $G_{xx}(r)=G_{yy}(r)$ displays long-range oscillatory behavior characteristic of AFM order, while $G_{zz}(r)$ decays rapidly (Fig.~S2 of the Supp.).

\subsection*{Pure ferromagnetic phase $(J_2=1)$}\label{4fm}

When only the FM interaction is present, the system stabilizes into the FM 4-sub phase upon cooling, illustrated in Fig.S1a of the Suppl. As shown in Fig.~\ref{fig:pure_fm}a, the orbital occupations become anisotropic below the transition temperature, with $\langle\tilde{n}_{zx}\rangle$ becoming larger than $\langle\tilde{n}_{xy}\rangle$ and $\langle\tilde{n}_{yz}\rangle$. The orbital imbalance reflects  spontaneous cubic symmetry breaking associated with the selection of a preferred magnetization direction. Assuming the magnetization lies along the $z$ axis, the resulting orbital configuration favors the $zx$ orbital while lifting the degeneracy among the three $t_{2g}$ orbitals.
The magnetic structure is characterized by a net moment along $z$, while the moments in the perpendicular plane form antiparallel pairs, with a $90\degree$ rotation between neighboring layers, giving rise to the characteristic four-sublattice arrangement. The onset of the order is captured by the staggered magnetization $m_{stg}^{xy}$ shown in Fig.~\ref{fig:pure_fm}b, which becomes finite below $T_c\sim0.8$. The transition coincides with the lifting of the orbital degeneracy, indicating the simultaneous development of the magnetic and orbital order.

As shown in Fig.~\ref{fig:pure_fm}c, the inverse susceptibility, \chipar, increases as the temperature is lowered, reflecting the suppression of fluctuations along the direction of the ordered moment. The \chiper remains comparatively small, corresponding to softer fluctuations within the perpendicular plane. These two are consistent with an FM ordering along the z direction. The inset in Fig.~\ref{fig:pure_fm}c further shows the deviation of $\chi^{-1}$ from the conventional Curie-Weiss (CW) behavior. Specifically, we find a negative CW temperature ($\Theta_{CW}$), consistent with experimental observations in Ba$_2$NaOsO$_6$~\cite{Erickson2007} and Ba$_2$MgReO$_6$~\cite{hirai2019successive}, both of which exhibit unconventional cAFM~\cite{Lu2017, Hirai2020}.

The real-space correlations further confirm the FM 4-sub structure as shown in Fig.~S3 of the Suppl. The correlation function $G_{zz}(r)$ saturates to a finite value at long distances, demonstrating long-range ferromagnetic order along the magnetization direction. In contrast, $G_{xx}(r)=G_{yy}(r)$ exhibit oscillatory behavior characteristic of the antiparallel in-plane arrangement between sublattices.

\subsection*{Pure quadrupolar phase $(V=1)$}
\begin{figure}[t]
    \centering
    \includegraphics[width=1\linewidth]{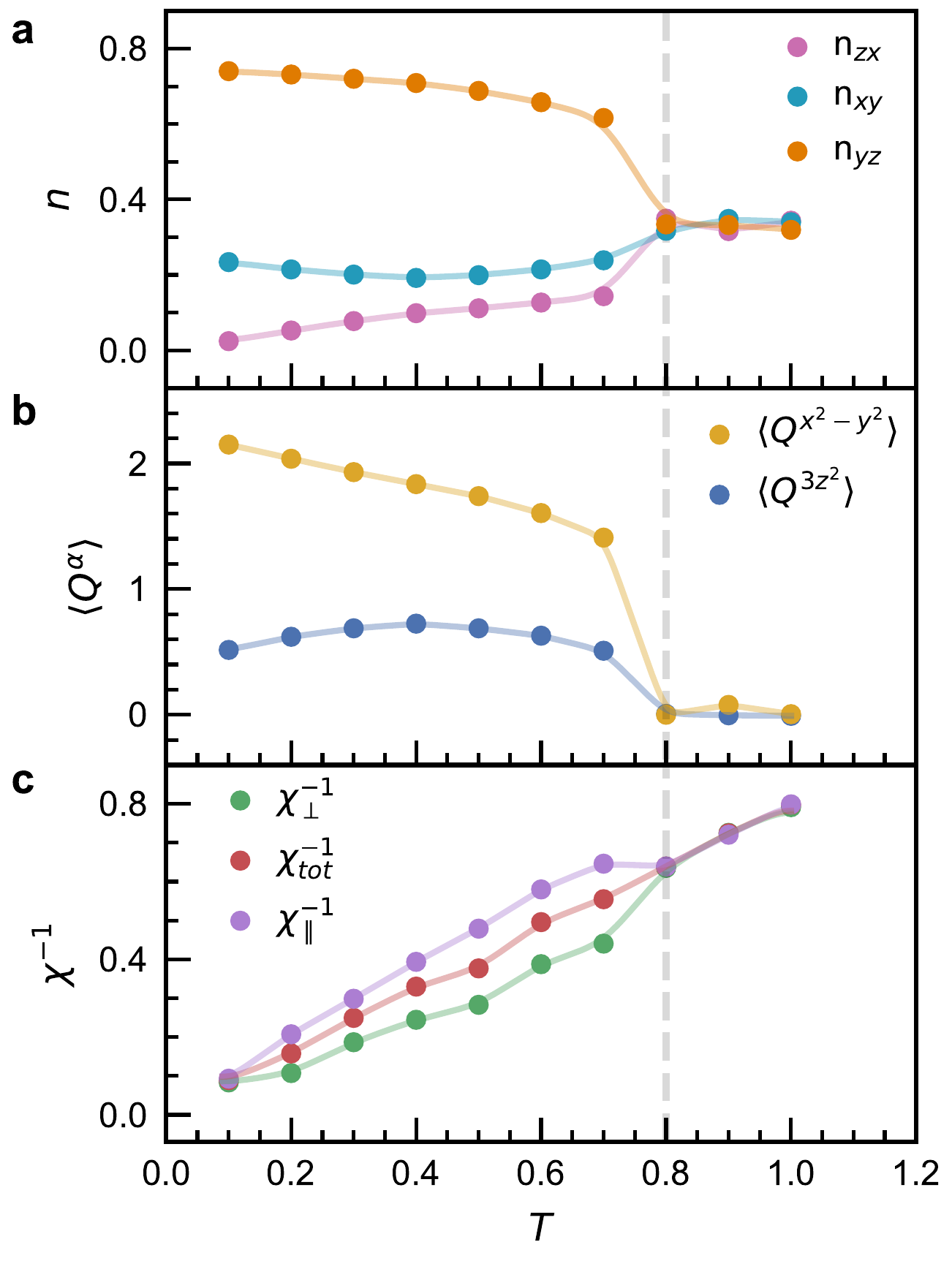}
    \caption{\textbf{Pure quadrupolar coupling.}
    \textbf{a} Orbital occupations $\langle\tilde{n}_{\alpha}\rangle$ vs. temperature, showing lifting of the degeneracy and the onset of quadrupolar order below $T_0 \sim 0.8$.
    \textbf{b} Quadrupolar order parameters $\langle Q^{3z^2}\rangle$ and $\langle Q^{x^2-y^2}\rangle$ developing below $T_0$ in the absence of dipolar magnetic order. 
    \textbf{c} Total ($\chi^{-1}$), parallel ($\chi_{\parallel}^{-1}$), and perpendicular ($\chi_{\perp}^{-1}$) inverse susceptibilities, remaining nearly isotropic across $T_0$ due to the lack of long-range magnetic order.  The absence of a pronounced anisotropy is consistent with the lack of long-range magnetic order in the quadrupolar phase.
    Gray dashed lines mark the transition temperature.    
    Results shown for $12^3\times4$ lattices. 
    }
    \label{fig:pure quad}
\end{figure}

In the pure quadrupolar limit, the system does not develop spontaneous magnetic ordering. Instead, the temperature evolution of the orbital occupations in Fig.~\ref{fig:pure quad}a shows that the three orbitals remain degenerate in the high-temperature phase. Below $T_0\sim0.8$, the orbital occupations separate, signaling the lifting of the orbital degeneracy and the onset of quadrupolar order. The orbital anisotropy reflects the spontaneous breaking of cubic symmetry through multipolar ordering rather than the magnetic one.

The development of quadrupolar order is captured by the quadrupolar moments $\langle Q^{3z^2}\rangle$ and $\langle Q^{x^2-y^2}\rangle$ shown in Fig.~\ref{fig:pure quad}b. As both quantities become finite below $T_0$ and the dipole moments remain zero, the phase represents pure quadrupolar order. The simultaneous presence of the two quadrupolar components is consistent with the reduced symmetry of the ordered state.
Unlike the two previous phases with dipolar magnetism, the inverse susceptibility, presented in Fig.~\ref{fig:pure quad}c, remains nearly isotropic, as \chipar and \chiper follow similar temperature dependence with a weak deviation from the PM behavior, consistent with the absence of an ordered dipole moment. Finally, the real-space correlation functions decay to zero at long distances in all directions, confirming that the transition is driven by quadrupolar rather than magnetic DOF (Fig.~S4 of the Suppl.).
\begin{figure*}[t]
    \centering
    \includegraphics[width=\linewidth]{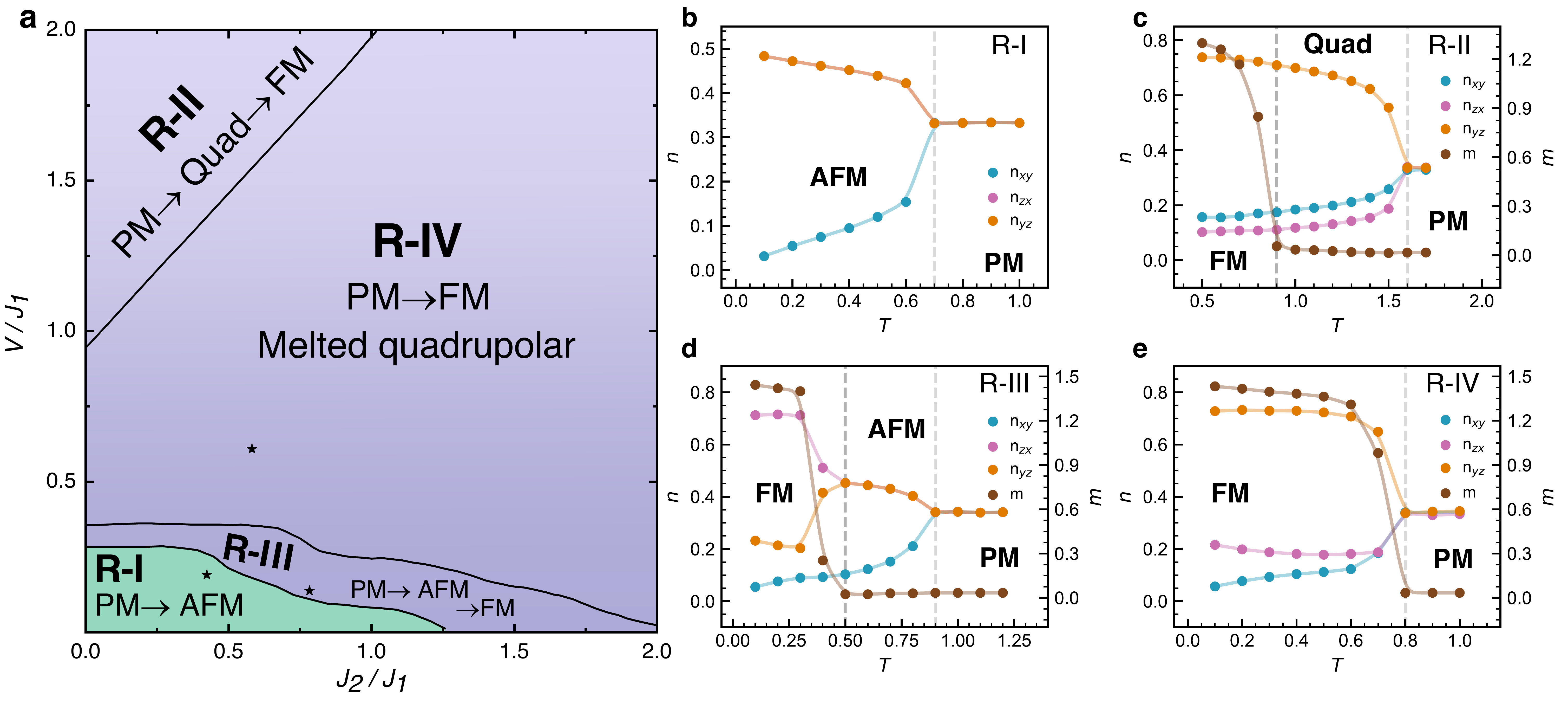}
    \caption{\textbf{Evolution of magnetic and quadrupolar phases.}
    \textbf{a} Phase diagram of the classical spin-orbit-coupled $d^1$ model versus $J_2/J_1$ and $V/J_1$ on a $12^3\times4$ lattice. The $T=0$ ground state comprises only the AFM 4-sub (shaded green) and FM 4-sub phases (shaded purple). Finite-temperature cooling sequences define four regions: I (PM $\rightarrow$ AFM 4-sub), II (PM $\rightarrow$ quadrupolar $\rightarrow$ FM 4-sub), III (PM $\rightarrow$ AFM 4-sub $\rightarrow$ FM 4-sub), and IV (direct PM $\rightarrow$ FM 4-sub). Stars mark parameter sets shown in Fig.~\ref{fig:comparison}.  
    Results shown for $12^3\times4$ lattices. 
    \textbf{b} Region I representative $T$-dependence, showing a single transition into the AFM 4-sub state. 
    \textbf{c} Region II evolution, where high-$T$ orbital splitting (quadrupolar order) precedes low-$T$ magnetization onset (FM 4-sub phase). 
    \textbf{d} Region III evolution, featuring intermediate AFM 4-sub order (orbital splitting, zero magnetization) before transitioning into the FM 4-sub ground state. 
    \textbf{e} Region IV evolution, exhibiting simultaneous onset of orbital polarization and net magnetization at a single $T_c$.
    }
    \label{fig:phase diagram}
\end{figure*}
\subsection*{The phase diagram}\label{phase diagram}
\begin{figure*}[t]
    \centering
    \includegraphics[width=\linewidth]{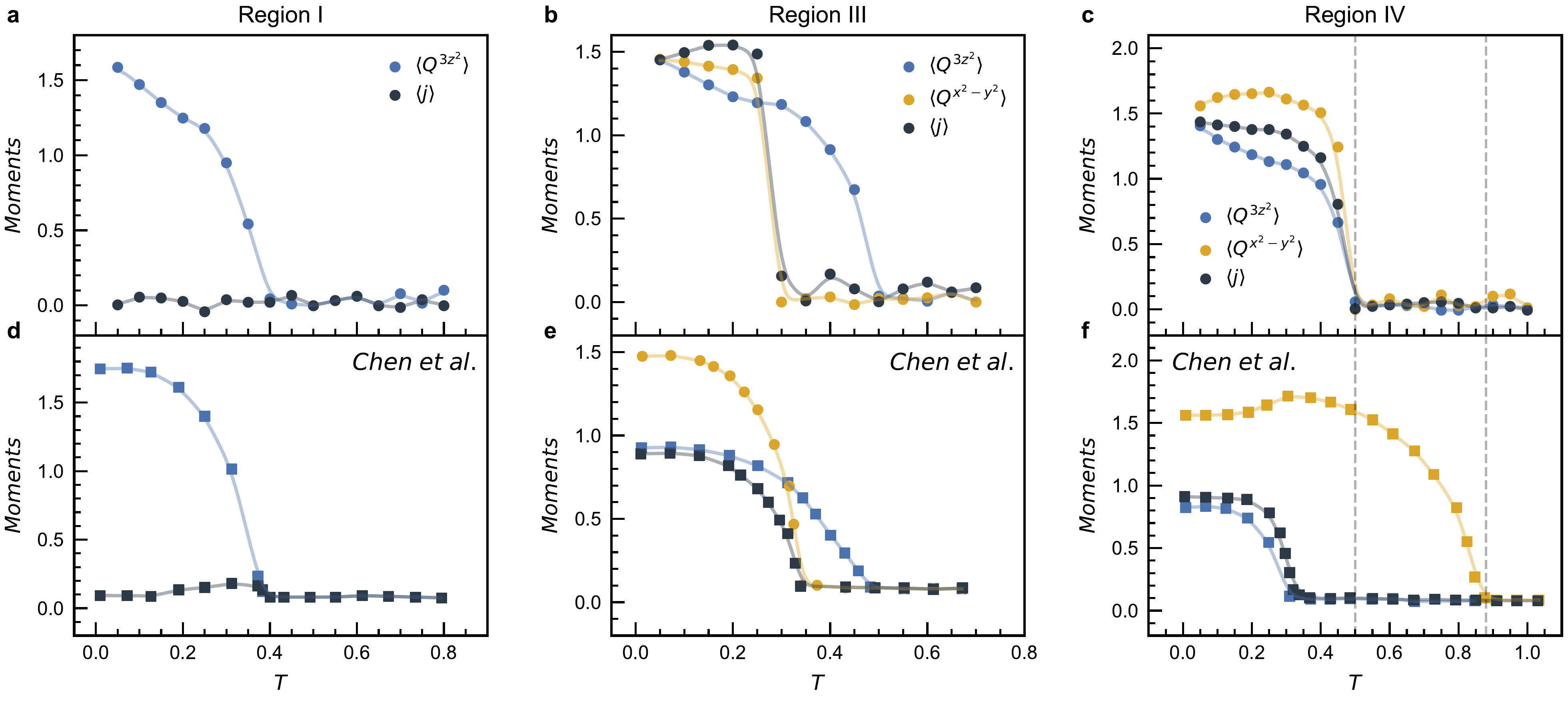}
    \caption{\textbf{Comparison between mean field~\cite{chen2010exotic} and Monte Carlo (this work).} Temperature dependence of order parameters across different phase regions. 
    \textbf{a, d} Region I (AFM 4-sub): Both methods show a direct PM-to-AFM transition with simultaneous magnetic and orbital ordering. 
    \textbf{b, e} Region III: Both calculations exhibit a two-step PM $\rightarrow$ quadrupolar $\rightarrow$ FM 4-sub transition upon cooling. 
    \textbf{c, f} Region IV: Unlike mean-field theory, which predicts an intermediate quadrupolar phase, MC yields a direct PM-to-FM transition due to thermal fluctuations, indicating the melting of the quadrupolar phase by thermal fluctuations.
    Dashed gray lines indicate the temperature range in which no intermediate quadrupolar phase is present in the MC results.
    The parameters used in each panel correspond to the starred points in Fig.~\ref{fig:phase diagram}. Curves in \textbf{d, e, f} have been reproduced, with permission, from~\cite{chen2010exotic}.}
    \label{fig:comparison}
\end{figure*}
Having characterized the isolated limits, we next explore how interaction competition governs the phase diagram when all Hamiltonian terms are active simultaneously. Specifically, we investigate the evolution of the magnetic and quadrupolar phases as a function of the interaction ratios: AFM over FM exchange ($J_2/J_1$) and electric quadrupole-quadrupolar over FM exchange ($V/J_1$). To obtain the detailed phase diagram, we perform a full Monte Carlo anneal from high to low temperature and identify the resulting zero-temperature ground state from magnetization, staggered magnetization, and quadrupolar moments.
The results for both zero- and finite-temperature are presented in Fig.~\ref{fig:phase diagram}a. At zero temperature, the phase diagram consists of only two ordered phases. For small values of both $J_2/J_1$ and $V/J_1$, the ground state is the AFM 4-sub phase. As either ratio is increased, the system transitions to the FM 4-sub phase.

To determine how the ground states are reached upon cooling, we additionally trace the temperature dependence of the orbital occupations and the net magnetization for representative points throughout the phase diagram. We find that the AFM and FM 4-sub ground states are approached through different sequences of phase transitions, giving rise to four distinct finite-temperature regions, illustrated by the representative temperature evolutions shown in Fig.~\ref{fig:phase diagram}b-e.
In Region I, corresponding to the AFM 4-sub ground state, the system undergoes a single transition directly from the PM phase into the AFM 4-sub state. As shown in Fig.~\ref{fig:phase diagram}b, the orbital occupations $n_{xy}$ and $n_{zx}(n_{yz})$ split below $T_c$, while the net magnetization remains zero throughout the transition, consistent with AFM ordering. Throughout the parameter range investigated, we do not observe any intermediate ordered phase preceding the AFM state.
Regions II, III, and IV all share the FM 4-sub ground state but differ in the intermediate phases encountered upon cooling. In Region II, shown in Fig.~\ref{fig:phase diagram}c, the orbital occupations split at the higher transition temperature $T_0$, whereas the net magnetization remains zero until the lower transition temperature $T_c$, where it becomes finite. This behavior identifies an intermediate quadrupolar phase preceding the FM 4-sub state. In Region III (Fig.~\ref{fig:phase diagram}d), the orbital occupations first split at $T_{c_2}$ while the net magnetization remains zero, indicating the onset of the AFM 4-sub phase. Upon further cooling to $T_{c_1}$, the orbital occupations evolve into the FM 4-sub configuration and the net magnetization simultaneously becomes finite, signaling the transition into the FM 4-sub ground state. Finally, in Region IV (Fig.~\ref{fig:phase diagram}e), the orbital occupations and the net magnetization develop simultaneously at a single transition temperature, indicating a direct transition from the PM phase into the FM 4-sub state without an intermediate ordered phase.

Beyond establishing the sequence of phase transitions, the phase diagram also reveals the distinct roles of the competing interactions in stabilizing the ordered states. In Region II, both transition temperatures, $T_0$ and \tc, scale linearly with the ferromagnetic exchange $J_2$ and remain nearly independent of the quadrupolar interaction $V$ (see Supplemental Material), indicating that the exchange interaction primarily controls the onset of both quadrupolar and magnetic order. In contrast, Region III displays a clear separation of energy scales. While the upper transition temperature $T_{c_2}$ follows the same $J_2$-driven behavior, the lower transition temperature $T_c$ depends strongly on $V$. This reflects the distinct quadrupolar structures of the intermediate AFM 4-sub and low-temperature FM 4-sub phases, making the quadrupolar interaction the key factor that stabilizes the transition between them.
 
Compared with the mean-field phase diagrams reported in Refs.~\cite{chen2010exotic,Svoboda2021}, the direct PM--FM 4-sub transition occupies a substantial portion of the ferromagnetic region where an intermediate quadrupolar phase was previously predicted. The microscopic origin of this suppression, together with the absence of the coplanar canted FM[110] state, is further discussed below.

\section*{Discussion}\label{discussion}

Our large-scale MC simulations demonstrate that a classical mapping of the effective moment $\mathbf{j}$ faithfully captures the overarching phase landscape predicted by prior quantum mean-field treatments~\cite{chen2010exotic, Svoboda2021}. This correspondence includes the stabilization of both the AFM 4-sub and FM 4-sub four-sublattice ground states, alongside the emergent intermediate-temperature quadrupolar phase. Such striking agreement reveals that the fundamental competition driving the phases is largely governed by classical interactions, remaining robust even when complex inter-site correlations beyond the mean-field approximation are explicitly included. At the same time, our calculations reveal two important differences with respect to the mean-field results~\cite{chen2010exotic, Svoboda2021}.
 
First, we establish the intrinsically quantum nature of the coplanar canted FM[110] state~\cite{chen2010exotic,Svoboda2021}. An FM moment along [110] can, in principle, arise from either a uniform alignment of local moments or a canted four-sublattice arrangement, although the former is energetically unfavorable. In the latter, moments on different layers point approximately along the $\langle100\rangle$ directions while producing a net magnetization along [110]. By contrast, our classical MC simulations produce only the FM 4-sub order throughout the FM region. While the single-occupancy constraint preserves the local orbital-occupation relation inherited from the quantum $j=3/2$ manifold, the classical degrees of freedom remain continuous vectors and therefore cannot represent the coherent superpositions of $m=\pm3/2\,,\pm1/2$ states that characterize the FM[110] wavefunction in the quantum mean-field description~\cite{chen2010exotic}. Consequently, the canted FM[110] state lies outside the classical model, indicating that its stabilization requires quantum degrees of freedom.

Second, we reveal the thermal-fluctuation-driven suppression of the intermediate quadrupolar phase. While mean-field theory predicts an extended quadrupolar regime across a broad parameter range~\cite{chen2010exotic, Svoboda2021}, our simulations show that spatial correlations reduce its stability and drive a direct transition into the FM 4-sub phase in part of the phase diagram. The temperature evolution of the dipolar and quadrupolar order parameters highlights this contrast between regimes (Fig.~\ref{fig:comparison}).
In Region III, quadrupolar moments emerge well above the magnetic ordering temperature, establishing a distinct intermediate quadrupolar phase before the system settles into a four-sublattice FM ground state (Fig.~\ref{fig:comparison}b). In Region IV, however, both order parameters develop simultaneously at a single transition temperature, completely erasing the intermediate phase (Fig.~\ref{fig:comparison}c). This represents a major departure from mean-field predictions, which widely overestimate the persistence of the preceding quadrupolar order over a broad parameter space. Thus, we underscore the vital role of spatial correlations and thermal fluctuations, which can only be captured by the simulation of expansive lattices.

Furthermore, we solidify the exotic dual-component quadrupolar phase characterized by the simultaneous coexistence of $\langle Q^{3z^2}\rangle$ and $\langle Q^{x^2-y^2}\rangle$ order parameters, only found in one of the previous studies. The co-emergence of both moments validates the four-sublattice framework of Svoboda\textit{ et al}.~\cite{Svoboda2021} while ruling out the restricted two-sublattice model of Chen \textit{et al}.~\cite{chen2010exotic}, where $\langle Q^{3z^2}\rangle$ is strictly zero. Our discovery demonstrates that the real-space geometry of the lattice unit cell dictates multipolar symmetry breaking. While standard two-sublattice descriptions artificially constrain the configurations, the four-sublattice architecture unlocks the vital structural freedom required to host complex, cooperative orbital ordering patterns.

Overall, our results demonstrate that a carefully constructed classical description can capture the essential finite-temperature physics of spin-orbit-entangled magnets. By explicitly incorporating long-range spatial correlations beyond the mean-field approximation, our simulations distinguish phenomena rooted in thermal fluctuations from those that require genuinely quantum DOF. Our work thus provides both physical insight into the origin of the competing ordered phases and an efficient, scalable framework for exploring complex multipolar phase diagrams beyond the reach of many fully quantum approaches.

\section*{Methods}
\label{sec:methods}
\noindent

\noindent{\bf Microscopic Model}\label{model app}\\
\begin{table*}[t]
\centering
\renewcommand{\arraystretch}{1.4}
\begin{tabular}{|c|c|c|c|} 
\hline
& \textbf{Chen \textit{et al.}} & \textbf{Present work} & \textbf{Svoboda \textit{et al.}} \\ 
\hline
Hamiltonian & $P_{3/2}\,H(J_{1}\,,J_2,\,V\,,\eta)\,P_{3/2}$ & $P_{3/2}H(J_{1}\,,J_2\,,V\,,\eta)P_{3/2}$ & \makecell{$H'(\lambda\,,J_{se}\,,V\,,\eta)=H(J_1\,,J_2\,,V\,,\eta)$\\ $+\,H_{AFM'}(J_{se}\,,\eta)$} \\ 
\hline
SOC & $\lambda \rightarrow \infty$ & $\lambda \rightarrow \infty$ & $\lambda$ finite\\ 
\hline
Basis of Hamiltonian & $j=3/2$ multiplets & $|j|=\sqrt{15}/2$ classical vector & $s=1/2$, $l_{\text{eff}}=-1$ \\ 
\hline
Method & Mean field & Classical Monte Carlo & Mean field\\ 
\hline
Number of sites/(unit cell) & 2 & $(6^3,8^3,10^3,12^3)\times 4$ & 4\\ 
\hline
Low temperature ground states & \makecell{FM[110]\\ AFM 2-sub} &\makecell{FM 4-sub\\ AFM 4-sub}&\makecell{Canted FM \\AFM 4-sub}\\ 
\hline
Intermediate temperature states & \makecell{Quadrupolar\\AFM 2-sub} & \makecell{Quadrupolar\\AFM 4-sub} & Quadrupolar\\ 
\hline
\end{tabular}
\caption{Comparison between the two mean field approaches~\cite{Svoboda2021, chen2010exotic} and the present work.}
\label{tab:1}
\end{table*}
The system is described by the total Hamiltonian introduced by Chen \textit{et al.}~\cite{chen2010exotic},
\begin{equation}
H=H_{ex-1}+H_{ex-2}+H_{quad}+H_{SO},
\label{tot hamiltonian}
\end{equation}
where $H_{ex-1}$ and $H_{ex-2}$ describe nearest-neighbor (NN) exchange interactions, $H_{quad}$ accounts for the electric quadrupole-quadrupole interaction, and $H_{SO}$ is the on-site SOC.
The first contribution describes the NN antiferromagnetic exchange arising from virtual electron hopping through oxygen $p$ orbitals,
\begin{equation}
H_{ex-1}=H^{XY}_{ex-1}+H^{YZ}_{ex-1}+H^{XZ}_{ex-1},
\end{equation}
where
\begin{equation}
H^{XY}_{ex-1}=J_{1}\sum_{\langle ij\rangle\in XY}\mathbf{S}_{i,xy}\cdot\mathbf{S}_{j,xy}
-\frac14n_{i,xy}n_{j,xy}.
\end{equation}
Here, $J_1>0$ is the AFM exchange coupling, while $\mathbf{S}_{i,xy}$ and $n_{i,xy}$ denote the spin and orbital occupation operators associated with the $xy$ orbital. The expressions for the $YZ$ and $XZ$ planes are obtained by cyclic permutation of the spatial indices.
The second contribution corresponds to the NN ferromagnetic exchange originating from virtual hopping between orthogonal orbitals,
\begin{equation}
H_{ex-2}=H^{XY}_{ex-2}+H^{YZ}_{ex-2}+H^{XZ}_{ex-2},
\end{equation}
with
\begin{equation}
\begin{split}
H^{XY}_{ex-2}=&-J_{2}\sum_{\langle ij\rangle\in XY}
\left[\mathbf{S}_{i,xy}\cdot(\mathbf{S}_{j,yz}+\mathbf{S}_{j,xz})
+\langle i\leftrightarrow j\rangle\right]\\[0.5em]
&+\frac32J'\sum_{\langle ij\rangle}n_{i,xy}n_{j,xy}.
\end{split}
\end{equation}
Here, $J_2>0$ is the FM exchange coupling.
The third contribution describes the electric quadrupole-quadrupole interaction,
\begin{equation}
H_{quad}=H^{XY}_{quad}+H^{YZ}_{quad}+H^{XZ}_{quad},
\end{equation}
where
\begin{equation}
\begin{split}
H^{XY}_{quad}=& \sum_{\langle ij\rangle\in XY} -\frac43V(n_{i,xz}-n_{i,yz})(n_{j,xz}-n_{j,yz})+\\
&+\frac{9V}{4}n_{i,xy}n_{j,xy}.
\end{split}
\end{equation}
The parameter $V$ denotes the quadrupolar coupling strength.
Finally, the on-site SOC is described by
\begin{equation}
H_{SO}=-\lambda\,\mathbf{L}\cdot\mathbf{S},
\end{equation}
where $\lambda$ is the SOC constant.

In the strong SOC limit, the Hamiltonian is projected onto the effective $j=3/2$ manifold~\cite{chen2010exotic}. Within the basis, the spin and orbital occupation operators become
\begin{equation}
\begin{split}
\tilde{S}^x_{i,xy}&=\frac14j^x_i-\frac13j^z_ij^x_ij^z_i,\\[0.5em]
\tilde{S}^y_{i,xy}&=\frac14j^y_i-\frac13j^z_ij^y_ij^z_i,\\[0.5em]
\tilde{S}^z_{i,xy}&=\frac14j^z_i-\frac13j^z_ij^z_ij^z_i,\\[0.5em]
\tilde{n}_{i,xy}&=\frac34-\frac13(j^z_i)^2,
\label{new operators}
\end{split}
\end{equation}
while the remaining operators are obtained by cyclic permutation of the spatial indices.
The projected orbital occupation operators satisfy the local single-occupancy $d^1$ constraint~\cite{chen2010exotic},
\begin{equation}
\begin{split}
\tilde n_{i,xy}+\tilde n_{i,yz}+\tilde n_{i,zx}=1,\\[0.5em]
0\le\tilde n_{i,xy},\tilde n_{i,yz},\tilde n_{i,zx}\le1,
\end{split}
\label{occupation_number1}
\end{equation}
which enforces the occupancy of a single $t_{2g}$ orbital at each lattice site. This constraint is preserved throughout the projection and plays a central role in the classical formulation of the model, as it directly relates the orbital occupations to the effective angular momentum DOF.
The Hamiltonian can therefore be rewritten in terms of the projected operators as
\begin{equation}
\tilde{H}=\tilde{H}_{ex-1}+\tilde{H}_{ex-2}+\tilde{H}_{quad}.
\end{equation}

\noindent{\bf Simulation protocol}\\
Equilibrium configurations at each temperature are obtained through simulated annealing. Starting from a random high-temperature configuration, the system is gradually cooled from $T_{max}=1–1.2$ to $T_{min}=0.05–0.1$, depending on the parameter set under consideration, using more than 10 linearly spaced temperature steps. At each temperature, 3000–5000 Metropolis sweeps are performed, depending on the system size, with the first $\sim 1000$ sweeps discarded for thermalization. All observables are averaged over the remaining sweeps.\\

\noindent{\bf Order parameters and observables}\label{model app}\\
To characterize the magnetic and orbital states, we compute several order parameters and observables. The net magnetization is defined as
\begin{equation}
m^\alpha = \left\langle \frac{1}{N}\sum_{R} j_R^\alpha \right\rangle\,,
\label{eq:magnetization}
\end{equation}
where $\langle...\rangle$ denotes the ensemble average.
To characterize magnetic states with staggered ordering within the xy plane, we also define the staggered magnetization
\begin{equation}
(m_{stg}^{(xy)})^2 = (m_{stg}^x)^2 + (m_{stg}^y)^2,
\label{eq:mstg_def}
\end{equation}
where 
\begin{equation}
\begin{split}
    (m_{stg}^{\alpha})^2 =& \Biggl\langle \frac{1}{N^2}\biggl[\left(\sum_{R}(-1)^{\sqrt{2}R_\alpha}\, j_R^{(\alpha)}\,\delta_{1,k}\right)^2 +\\
    + & \left(\sum_{R}(-1)^{\sqrt{2}R_\alpha}\, j_R^{(\beta)}\,\delta_{-1,k}\right)^2\biggr]\Biggr\rangle
\end{split}
\label{eq:mstg}
\end{equation}
with $\alpha\,,\beta=x\,,y$ and $k=(-1)^{\sqrt{2}R_z}$. This quantity is used to detect states in which the magnetic moments are staggered between neighboring layers while retaining an in-plane ordering pattern.
We then compute the magnetic susceptibility,
\begin{equation}
\chi_{\alpha\alpha} = \frac{N}{T}\left(\left\langle (j_R^{(\alpha)})^2\right\rangle - \left\langle j_R^{(\alpha)}\right\rangle^2\right)\,,
\label{eq:chi}
\end{equation}
with its longitudinal component along $z$, transverse component in the $xy$ plane, and the averaged susceptibility obtained from all three directions combined,
\begin{equation}
\begin{split}
    &\chi_\parallel = \chi_{zz}\,,\\
    &\chi_\perp = \frac{1}{2}(\chi_{xx}+\chi_{yy})\,,\\
    &\chi = \frac{1}{3}(\chi_{xx}+\chi_{yy}+\chi_{zz})\,.
\end{split}
\label{eq:chi_components}
\end{equation}
Here, the labels longitudinal and transverse refer to the crystalline $z$ axis and $xy$ plane, respectively, rather than to the direction of the ordered moment.
Such a decomposition allows us to distinguish fluctuations along the $z$ axis from those within the $xy$ plane.
We additionally compute the real-space spin-spin correlation function along each direction
\begin{equation}
G_{\alpha\alpha}(r) = \left\langle j^{(\alpha)}(r)\cdot j^{(\alpha)}(0)\right\rangle\,,
\label{eq:correlation}
\end{equation}
as a function of the distance $r$ between sites. A correlation function that decays to zero at large $r$ indicates the absence of long-range order in that direction, while a finite asymptotic value or a non-decaying oscillatory behavior signals long-range FM or AFM order, respectively.
Finally, since the orbital DOF in this model carry an electric quadrupole moment, we characterize the quadrupolar order through
\begin{equation}
Q^{3z^2}_i = \frac{3(j^z_i)^2 - j_i^2}{\sqrt{3}}\,,\qquad Q^{x^2-y^2}_i = (j^x_i)^2 - (j^y_i)^2\,,
\label{eq:quadrupole}
\end{equation}
These quantities measure the orbital polarization associated with the quadrupolar DOF. Specifically, $Q^{3z^2}$ describes the relative occupation of the $xy$ orbital with respect to the $yz/zx$ orbitals, while $Q^{x^2-y^2}$ distinguishes between the $yz$ and $zx$ orbital occupations.

\section*{Data availability}
All data supporting the findings of this study are available within the paper and its Supplementary Information.

\section*{Acknowledgments}
\label{Ack}
The authors thank Hitesh Changlani, Yiou Zhang, Ravindra Nanguneri and Kemp Plumb for useful discussions. 

\section*{Funding}
This work was supported in part by U.S. National Science Foundation grant DMR-1905532 (V.F.M.). The calculations presented here were performed using resources at the Center for Computation and Visualization, Brown University, which is supported by NSF Grant No. ACI-1548562. 

\section*{Competing interests}
The authors declare no competing interests.

\section*{Author contributions}
R.C. performed the modeling and simulations.
G.C. led the interpretation of the results. 
G.C., I.K.N. and R.C. wrote the manuscript. 
W.Z. discussed the results.
N.T. and V.F.M. conceptualized and guided the work.
All authors reviewed the manuscript.

\section*{Additional information}
Correspondence should be addressed to Vesna F. Mitrovi\'c.

\bibliography{biblio}{}

\clearpage
\appendix 

\begin{widetext}
\title{Thermally melted quadrupolar order and intrinsically quantum phases in 5$d^1$ double perovskites
  ${}$\\
-- Supplemental Material --}

\author{Rong Cong}
\thanks{Current affiliation: NHMFL, FL}
\thanks{Equal contribution}
\affiliation{Department of Physics, Brown University, Providence, Rhode Island 02912, USA}
\author{Ginevra Corsale}
\thanks{Equal contribution}
\affiliation{Department of Physics, Brown University, Providence, Rhode Island 02912, USA}
\author{Ilija K. Nikolov}
\affiliation{Department of Physics, Brown University, Providence, Rhode Island 02912, USA}
\author{Wenjuan Zhang}
\affiliation{Department of Physics, Ohio State University, 191 West Woodruff Ave
Columbus, Ohio 43210, USA}
\author{Nandini Trivedi}
\affiliation{Department of Physics, Ohio State University, 191 West Woodruff Ave
Columbus, Ohio 43210, USA}
\author{Vesna F.  Mitrovi\'c}
\affiliation{Department of Physics, Brown University, Providence, Rhode Island 02912, USA}
\affiliation{Brown Center for Theoretical Physics and Innovation, BCTPI, Brown University, Providence, Rhode Island 02912-1843, USA}

\maketitle
\tableofcontents
\clearpage

\setcounter{section}{0}
\setcounter{equation}{0}
\setcounter{figure}{0}
\setcounter{table}{0}
\setcounter{page}{1}
\makeatletter

\renewcommand{\thefigure}{S\arabic{figure}}
\section{Magnetic and orbital ordering}
\begin{figure}[H]
    \centering
    \includegraphics[width=0.35\linewidth]{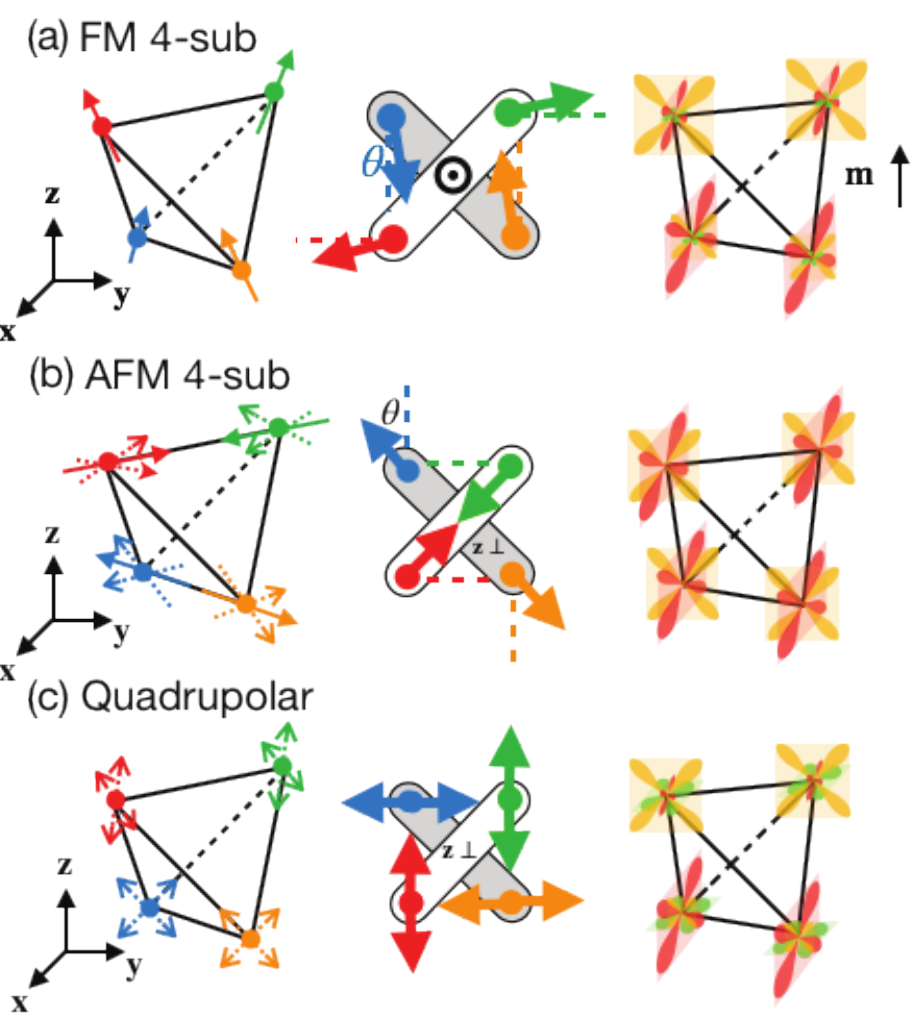}
    \caption{\textbf{Magnetic structures and orbital occupations of the ordered phases.}
    \textbf{a} FM 4-sub phase, \textbf{b} AFM 4-sub phase, and \textbf{c} quadrupolar phase.
    The first column illustrates the orientations of the effective moments on the four sites of a tetrahedron, corresponding to the four sublattices of the FCC structure. The dotted multiple arrows indicate that the local moments fluctuate among the symmetry-equivalent directions compatible with the corresponding ordered state. The second column shows the same configurations projected onto the $xy$ plane, viewed along the $z$ axis. Moments belonging to the same layer are connected by diagonal lines to emphasize the four-sublattice arrangement. The third column illustrates the corresponding orbital occupations on each site. The $\tilde n_{xy}$, $ \tilde n_{yz}$, and $\tilde n_{zx}$ orbitals are represented by green, yellow, and red lobes, respectively, with the lobe sizes qualitatively indicating their relative occupancies. In the FM 4-sub and AFM 4-sub phases, the orbital occupations become anisotropic, reflecting the spontaneous breaking of cubic symmetry. In contrast, the quadrupolar phase exhibits no dipolar magnetic order, while the orbital occupations remain anisotropic owing to the development of quadrupolar order.}
    \label{fig:sup_orb}
\end{figure}

\clearpage
\section{Correlation functions}
\begin{figure}[H]
    \centering
    \includegraphics[width=.65\linewidth]{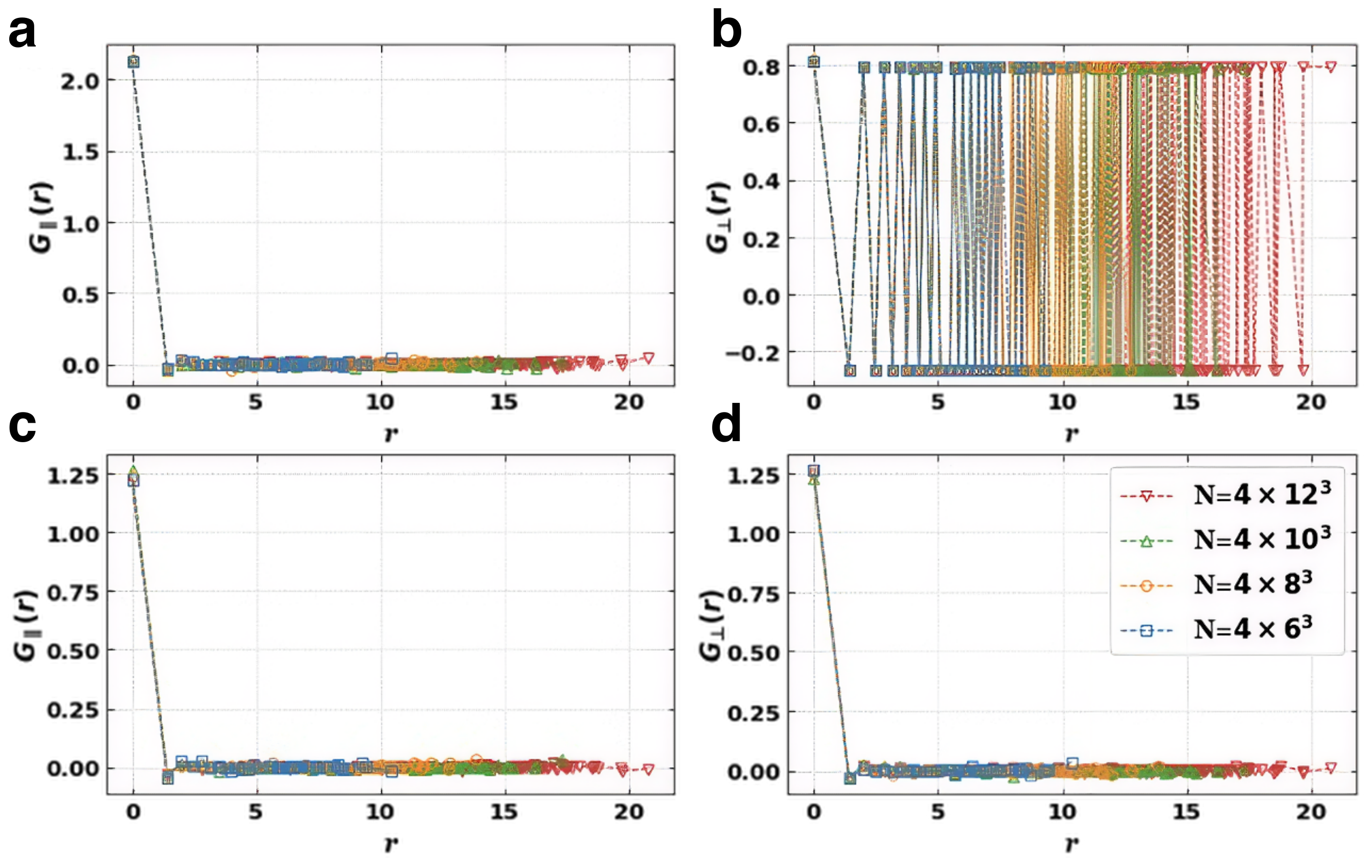}
    \caption{\textbf{Real-space correlation functions for the pure AFM 4-sub phase.}
    \textbf{a}, \textbf{b} Longitudinal, perpendicular components of the correlations in the ordered phase ($T<T_c$). Below $T_c$, the in-plane correlations display the oscillatory behavior characteristic of the AFM 4-sub ordering, while the longitudinal correlations remain short-ranged, consistent with moments ordered within the $xy$ plane.
    \textbf{c}, \textbf{d} The lower panels show the same quantities but above the transition ($T>T_c$), where both longitudinal and transverse correlations decay to zero at long distances, indicating the absence of long-range magnetic order.
   }
    \label{fig:supp_afm}
\end{figure}

\begin{figure}[H]
    \centering
    \includegraphics[width=.65\linewidth]{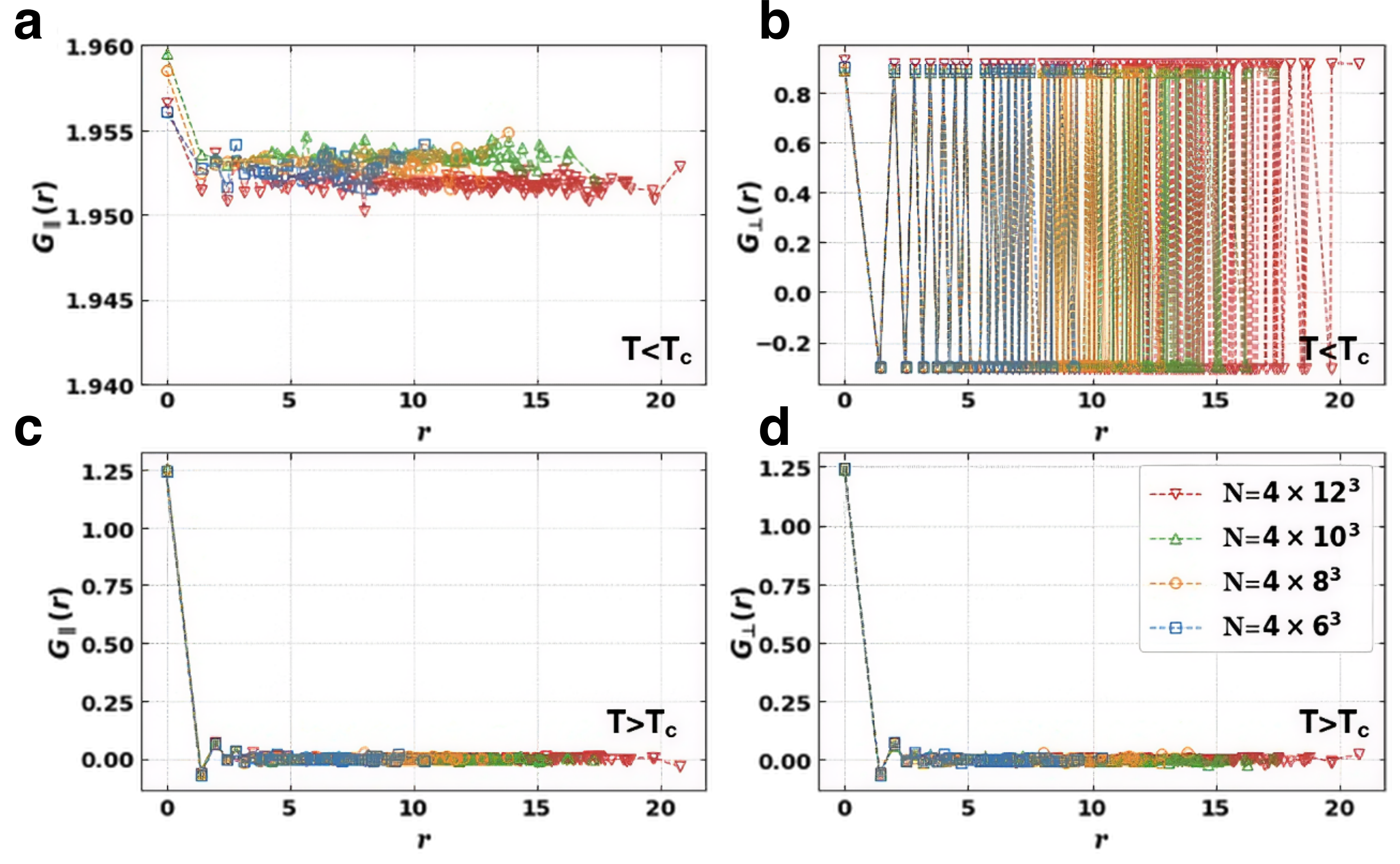}
    \caption{\textbf{Real-space correlation functions for the pure FM 4-sub phase.}
    \textbf{a}, \textbf{b} Longitudinal, perpendicular components The of the correlations in the ordered phase ($T<T_c$). Below $T_c$, the longitudinal correlation function saturates at long distances, indicating long-range ferromagnetic order along the magnetization direction, while the perpendicular correlations exhibit oscillatory behavior associated with the staggered in-plane arrangement of the four-sublattice structure.
    \textbf{c}, \textbf{d} The lower panels show the correlations above the transition ($T>T_c$), where both longitudinal and transverse correlations decay to zero at long distances, indicating the loss of long-range magnetic order.
    }
    \label{fig:supp_fm}
\end{figure}

\begin{figure}[H]
    \centering
    \includegraphics[width=.65\linewidth]{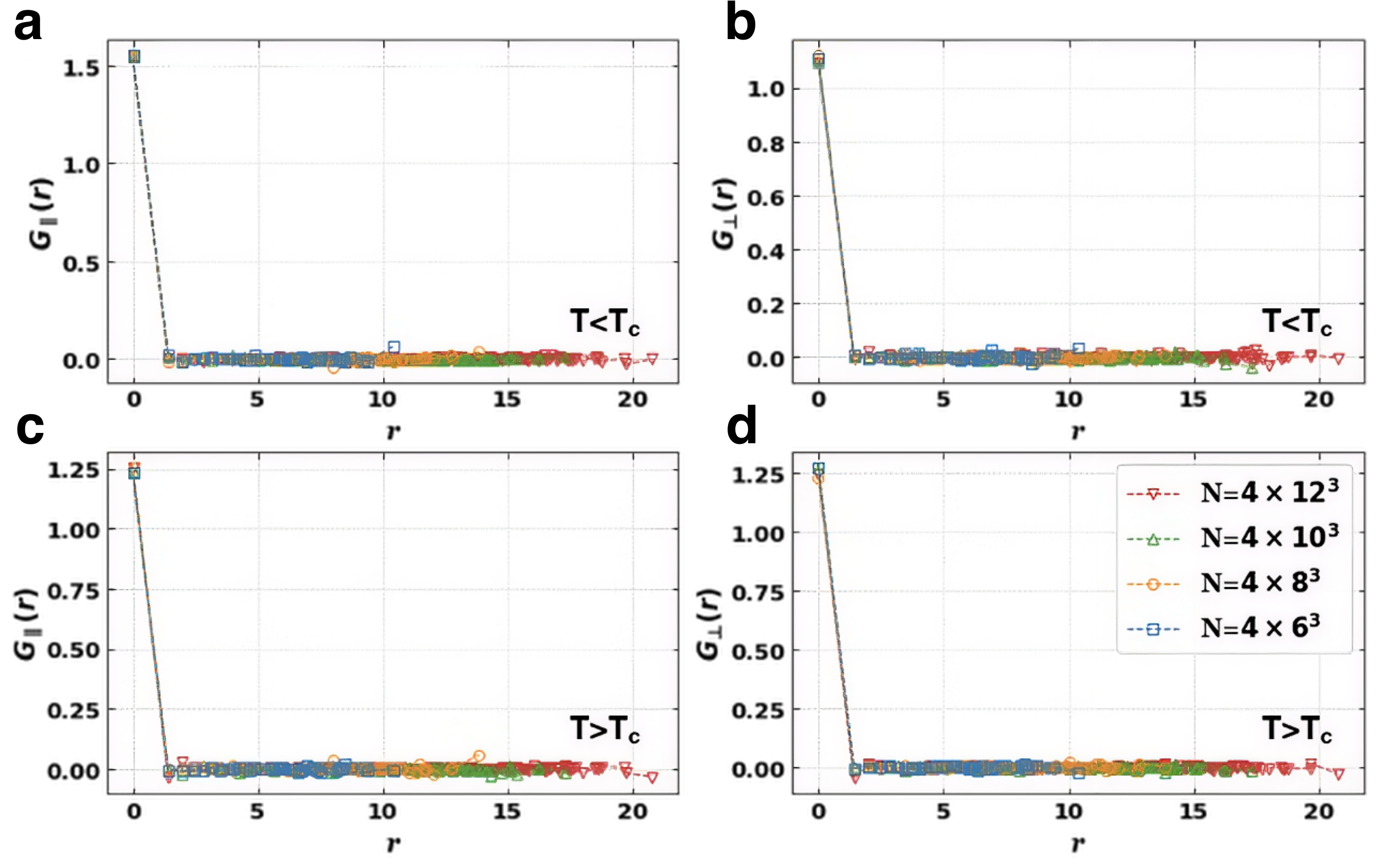}
    \caption{\textbf{Real-space magnetic correlation functions for the pure quadrupolar phase.}
    \textbf{a}, \textbf{b} The upper panels show the correlations in the ordered quadrupolar phase ($T<T_0$), corresponding to the longitudinal and perpendicular components, respectively.
    \textbf{c}, \textbf{d} The lower panels show the correlations above the transition ($T>T_0$).
    In both temperature regimes, the longitudinal and transverse magnetic correlations decay to zero at long distances, confirming the absence of long-range dipolar order in the quadrupolar phase.}
    \label{fig:supp_quad}
\end{figure}

\section{Transition temperatures evolution}

\begin{figure}[H]
    \centering
    \includegraphics[width=0.7\linewidth]{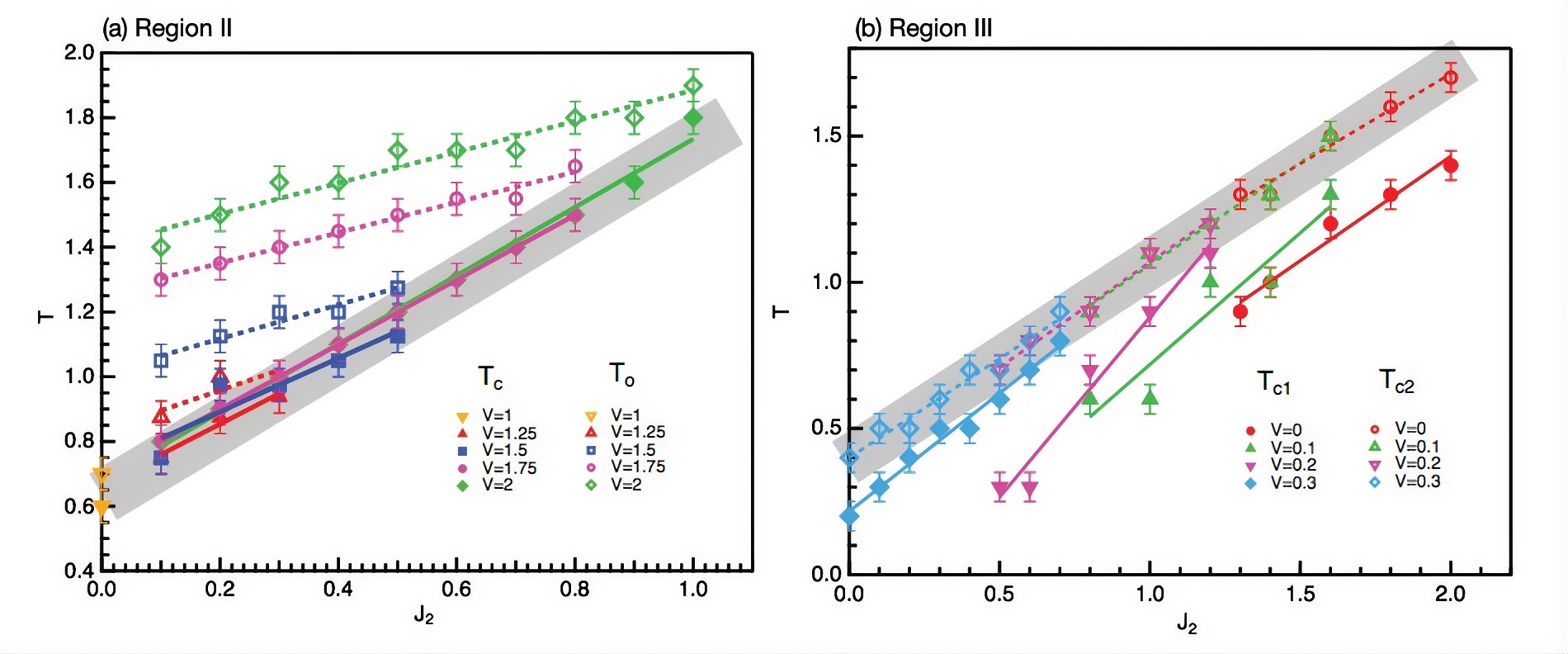}
    \caption{\textbf{Successive transition temperatures in Regions II and III.}
    Transition temperatures as a function of the ferromagnetic exchange coupling $J_2$ for several values of the quadrupolar interaction $V$~($J_1=1$). \textbf{a} Transition temperatures for Region II. The quadrupolar transition temperature $T_0$ (PM$\rightarrow$QP) and the magnetic transition temperature $T_c$ (QP$\rightarrow$FM 4-sub) both increase approximately linearly with $J_2$. The slopes are nearly independent of $V$, indicating that the quadrupolar interaction has little influence on the evolution of either transition. \textbf{b} Transition temperatures for Region III. The upper transition temperature $T_{c_2}$ (PM$\rightarrow$AFM 4-sub) likewise increases linearly with $J_2$ and is nearly independent of $V$. In contrast, the lower transition temperature $T_{c_1}$ (AFM 4-sub$\rightarrow$FM 4-sub) shifts with increasing $V$. This distinct behavior originates from the different quadrupolar order associated with the AFM 4-sub and FM 4-sub phases. Consequently, the quadrupolar interaction contributes to the energetics of the lower transition, whereas the upper transition is primarily governed by the exchange interaction $J_2$. Error bars represent the spread in transition temperatures obtained for different system sizes, and the gray shaded regions serve as guides to the eye.}
    \label{fig:supp_temp}
\end{figure}

\section{Finite-size dependence}
To assess the effect of the system size on the finite-temperature behavior, we repeated the simulations for lattice sizes ranging from $6^3\times4$ to $12^3\times4$. Figure~\ref{fig:size_effect} shows the temperature dependence of the relevant order parameters for representative parameter sets belonging to the four regions of the phase diagram.
The overall behavior of the curves is found to be independent of the system size considered. The main difference is observed for the smallest system, $6^3\times4$, where the transitions are slightly broadened due to finite-size effects. As the system size increases, the transition features become progressively sharper, while the transition temperatures and the sequence of phases remain essentially unchanged. In particular, the results obtained for the $8^3\times4$, $10^3\times4$, and $12^3\times4$ lattices are nearly indistinguishable, indicating that the largest system size used in our simulations is sufficient to capture the thermodynamic behavior of the model.
\begin{figure}[h]
    \centering
    \includegraphics[width=0.9\linewidth]{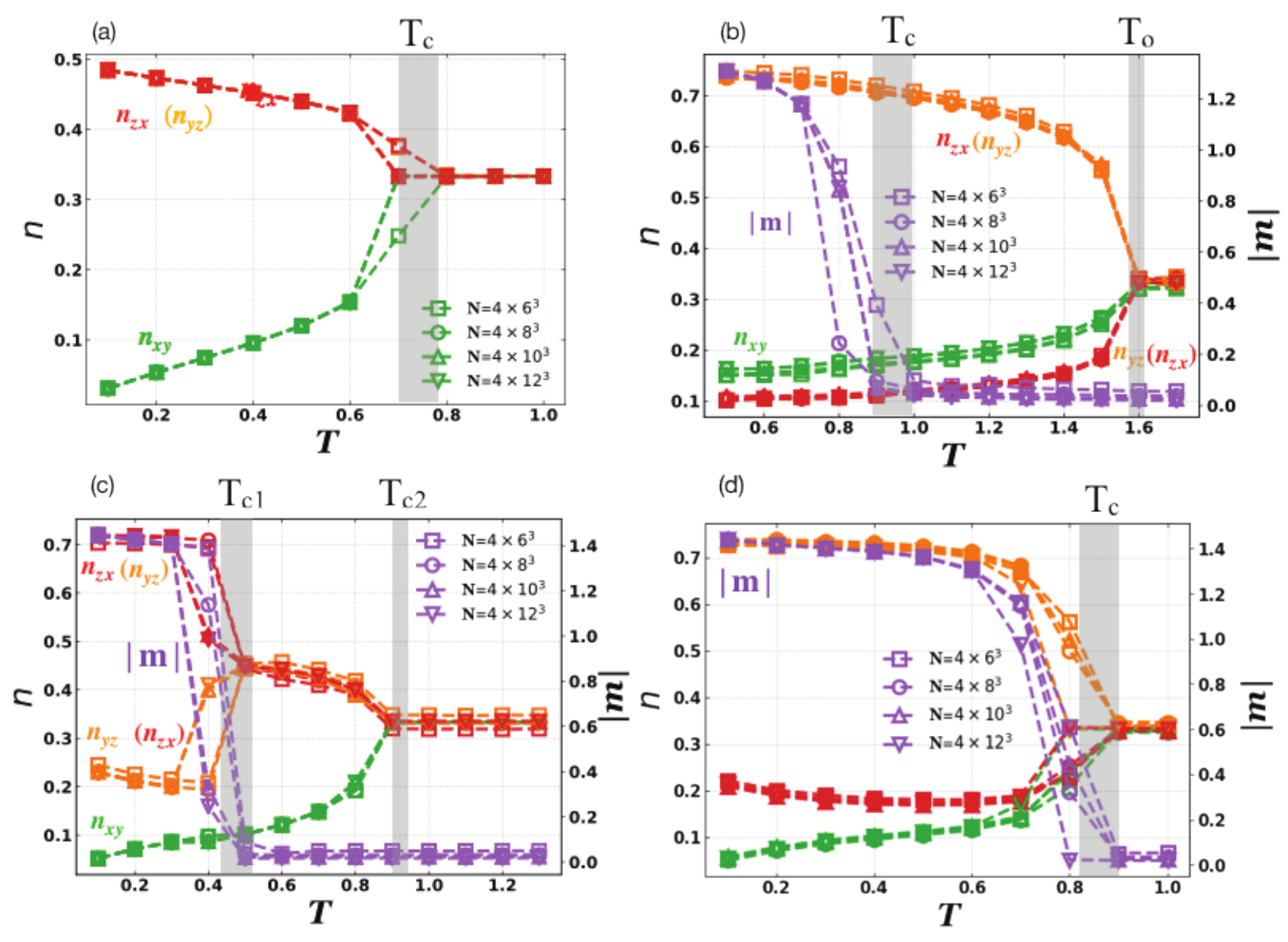}
    \caption{\textbf{Finite-size dependence of the thermal evolution in different regions of the phase diagram.}
    Temperature dependence of the relevant order parameters for representative parameter sets in Regions I-IV, obtained for lattice sizes $L^3\times4$ with $L=6,8,10,$ and $12$. The qualitative behavior of the thermal transitions is unchanged with increasing system size. The main finite-size effect is a progressive sharpening of the transition features, while the transition temperatures remain essentially unchanged for $L\geq8$.}
    \label{fig:size_effect}
\end{figure} 
\end{widetext}

\end{document}